\documentclass[reprint,amsmath,amssymb,aps,floatfix,superscriptaddress,showpacs,prl]{revtex4-1}

\usepackage{graphicx}% Include figure files
\usepackage{dcolumn}% Align table columns on decimal point
\usepackage{bm}% bold math% You should use BibTeX and apsrev.bst for references
\usepackage{color}
\usepackage[inline]{enumitem} %used for inline enumerated lists
\usepackage{url}

\begin{document}
{

\title{Direct Measurement of the Cosmic-Ray Helium Spectrum\\ 
 from 40~GeV to 250~TeV with the Calorimetric Electron Telescope\\
 on the International Space Station}

% Updated on 2023.01.14
%
\author{O.~Adriani}
\affiliation{Department of Physics, University of Florence, Via Sansone, 1 - 50019, Sesto Fiorentino, Italy}
\affiliation{INFN Sezione di Firenze, Via Sansone, 1 - 50019, Sesto Fiorentino, Italy}
\author{Y.~Akaike}
\affiliation{Waseda Research Institute for Science and Engineering, Waseda University, 17 Kikuicho,  Shinjuku, Tokyo 162-0044, Japan}
\affiliation{JEM Utilization Center, Human Spaceflight Technology Directorate, Japan Aerospace Exploration Agency, 2-1-1 Sengen, Tsukuba, Ibaraki 305-8505, Japan}
\author{K.~Asano}
\affiliation{Institute for Cosmic Ray Research, The University of Tokyo, 5-1-5 Kashiwa-no-Ha, Kashiwa, Chiba 277-8582, Japan}
\author{Y.~Asaoka}
\affiliation{Institute for Cosmic Ray Research, The University of Tokyo, 5-1-5 Kashiwa-no-Ha, Kashiwa, Chiba 277-8582, Japan}
\author{E.~Berti} 
%\affiliation{Department of Physics, University of Florence, Via Sansone, 1 - 50019, Sesto Fiorentino, Italy}
\affiliation{INFN Sezione di Firenze, Via Sansone, 1 - 50019, Sesto Fiorentino, Italy}
\affiliation{Institute of Applied Physics (IFAC),  National Research Council (CNR), Via Madonna del Piano, 10, 50019, Sesto Fiorentino, Italy}\author{G.~Bigongiari}
\affiliation{Department of Physical Sciences, Earth and Environment, University of Siena, via Roma 56, 53100 Siena, Italy}
\affiliation{INFN Sezione di Pisa, Polo Fibonacci, Largo B. Pontecorvo, 3 - 56127 Pisa, Italy}
\author{W.R.~Binns}
\affiliation{Department of Physics and McDonnell Center for the Space Sciences, Washington University, One Brookings Drive, St. Louis, Missouri 63130-4899, USA}
\author{M.~Bongi}
\affiliation{Department of Physics, University of Florence, Via Sansone, 1 - 50019, Sesto Fiorentino, Italy}
\affiliation{INFN Sezione di Firenze, Via Sansone, 1 - 50019, Sesto Fiorentino, Italy}
\author{P.~Brogi}
\email{paolo.brogi@unisi.it}
\affiliation{Department of Physical Sciences, Earth and Environment, University of Siena, via Roma 56, 53100 Siena, Italy}
\affiliation{INFN Sezione di Pisa, Polo Fibonacci, Largo B. Pontecorvo, 3 - 56127 Pisa, Italy}
\author{A.~Bruno}
\affiliation{Heliospheric Physics Laboratory, NASA/GSFC, Greenbelt, Maryland 20771, USA}
\author{J.H.~Buckley}
\affiliation{Department of Physics and McDonnell Center for the Space Sciences, Washington University, One Brookings Drive, St. Louis, Missouri 63130-4899, USA}
\author{N.~Cannady}
\affiliation{Center for Space Sciences and Technology, University of Maryland, Baltimore County, 1000 Hilltop Circle, Baltimore, Maryland 21250, USA}
\affiliation{Astroparticle Physics Laboratory, NASA/GSFC, Greenbelt, Maryland 20771, USA}
\affiliation{Center for Research and Exploration in Space Sciences and Technology, NASA/GSFC, Greenbelt, Maryland 20771, USA}
\author{G.~Castellini}
\affiliation{Institute of Applied Physics (IFAC),  National Research Council (CNR), Via Madonna del Piano, 10, 50019, Sesto Fiorentino, Italy}
\author{C.~Checchia}
\affiliation{Department of Physical Sciences, Earth and Environment, University of Siena, via Roma 56, 53100 Siena, Italy}
\affiliation{INFN Sezione di Pisa, Polo Fibonacci, Largo B. Pontecorvo, 3 - 56127 Pisa, Italy}
\author{M.L.~Cherry}
\affiliation{Department of Physics and Astronomy, Louisiana State University, 202 Nicholson Hall, Baton Rouge, Louisiana 70803, USA}
\author{G.~Collazuol}
\affiliation{Department of Physics and Astronomy, University of Padova, Via Marzolo, 8, 35131 Padova, Italy}
\affiliation{INFN Sezione di Padova, Via Marzolo, 8, 35131 Padova, Italy} 
\author{G.A.~de~Nolfo}
\affiliation{Heliospheric Physics Laboratory, NASA/GSFC, Greenbelt, Maryland 20771, USA}
\author{K.~Ebisawa}
\affiliation{Institute of Space and Astronautical Science, Japan Aerospace Exploration Agency, 3-1-1 Yoshinodai, Chuo, Sagamihara, Kanagawa 252-5210, Japan}
\author{A.~W.~Ficklin}
\affiliation{Department of Physics and Astronomy, Louisiana State University, 202 Nicholson Hall, Baton Rouge, Louisiana 70803, USA}
\author{H.~Fuke}
\affiliation{Institute of Space and Astronautical Science, Japan Aerospace Exploration Agency, 3-1-1 Yoshinodai, Chuo, Sagamihara, Kanagawa 252-5210, Japan}
\author{S.~Gonzi}
\affiliation{Department of Physics, University of Florence, Via Sansone, 1 - 50019, Sesto Fiorentino, Italy}
\affiliation{INFN Sezione di Firenze, Via Sansone, 1 - 50019, Sesto Fiorentino, Italy}
\affiliation{Institute of Applied Physics (IFAC),  National Research Council (CNR), Via Madonna del Piano, 10, 50019, Sesto Fiorentino, Italy}\author{T.G.~Guzik}
\affiliation{Department of Physics and Astronomy, Louisiana State University, 202 Nicholson Hall, Baton Rouge, Louisiana 70803, USA}
\author{T.~Hams}
\affiliation{Center for Space Sciences and Technology, University of Maryland, Baltimore County, 1000 Hilltop Circle, Baltimore, Maryland 21250, USA}
\author{K.~Hibino}
\affiliation{Kanagawa University, 3-27-1 Rokkakubashi, Kanagawa, Yokohama, Kanagawa 221-8686, Japan}
\author{M.~Ichimura}
\affiliation{Faculty of Science and Technology, Graduate School of Science and Technology, Hirosaki University, 3, Bunkyo, Hirosaki, Aomori 036-8561, Japan}
\author{K.~Ioka}
\affiliation{Yukawa Institute for Theoretical Physics, Kyoto University, Kitashirakawa Oiwake-cho, Sakyo-ku, Kyoto, 606-8502, Japan}
\author{W.~Ishizaki}
\affiliation{Institute for Cosmic Ray Research, The University of Tokyo, 5-1-5 Kashiwa-no-Ha, Kashiwa, Chiba 277-8582, Japan}
\author{M.H.~Israel}
\affiliation{Department of Physics and McDonnell Center for the Space Sciences, Washington University, One Brookings Drive, St. Louis, Missouri 63130-4899, USA}
\author{K.~Kasahara}
\affiliation{Department of Electronic Information Systems, Shibaura Institute of Technology, 307 Fukasaku, Minuma, Saitama 337-8570, Japan}
\author{J.~Kataoka}
\affiliation{School of Advanced Science and	Engineering, Waseda University, 3-4-1 Okubo, Shinjuku, Tokyo 169-8555, Japan}
\author{R.~Kataoka}
\affiliation{National Institute of Polar Research, 10-3, Midori-cho, Tachikawa, Tokyo 190-8518, Japan}
\author{Y.~Katayose}
\affiliation{Faculty of Engineering, Division of Intelligent Systems Engineering, Yokohama National University, 79-5 Tokiwadai, Hodogaya, Yokohama 240-8501, Japan}
\author{C.~Kato}
\affiliation{Faculty of Science, Shinshu University, 3-1-1 Asahi, Matsumoto, Nagano 390-8621, Japan}
\author{N.~Kawanaka}
\affiliation{Yukawa Institute for Theoretical Physics, Kyoto University, Kitashirakawa Oiwake-cho, Sakyo-ku, Kyoto, 606-8502, Japan}
\author{Y.~Kawakubo}
\affiliation{Department of Physics and Astronomy, Louisiana State University, 202 Nicholson Hall, Baton Rouge, Louisiana 70803, USA}
\author{K.~Kobayashi}
\email{kenkou@aoni.waseda.jp}
\affiliation{Waseda Research Institute for Science and Engineering, Waseda University, 17 Kikuicho,  Shinjuku, Tokyo 162-0044, Japan}
\affiliation{JEM Utilization Center, Human Spaceflight Technology Directorate, Japan Aerospace Exploration Agency, 2-1-1 Sengen, Tsukuba, Ibaraki 305-8505, Japan}
\author{K.~Kohri} 
\affiliation{Institute of Particle and Nuclear Studies, High Energy Accelerator Research Organization, 1-1 Oho, Tsukuba, Ibaraki, 305-0801, Japan} 
\author{H.S.~Krawczynski}
\affiliation{Department of Physics and McDonnell Center for the Space Sciences, Washington University, One Brookings Drive, St. Louis, Missouri 63130-4899, USA}
\author{J.F.~Krizmanic}
\affiliation{Astroparticle Physics Laboratory, NASA/GSFC, Greenbelt, Maryland 20771, USA}
\author{P.~Maestro}
\affiliation{Department of Physical Sciences, Earth and Environment, University of Siena, via Roma 56, 53100 Siena, Italy}
\affiliation{INFN Sezione di Pisa, Polo Fibonacci, Largo B. Pontecorvo, 3 - 56127 Pisa, Italy}
\author{P.S.~Marrocchesi}
\affiliation{Department of Physical Sciences, Earth and Environment, University of Siena, via Roma 56, 53100 Siena, Italy}
\affiliation{INFN Sezione di Pisa, Polo Fibonacci, Largo B. Pontecorvo, 3 - 56127 Pisa, Italy}
\author{A.M.~Messineo}
\affiliation{INFN Sezione di Pisa, Polo Fibonacci, Largo B. Pontecorvo, 3 - 56127 Pisa, Italy}
\affiliation{University of Pisa, Polo Fibonacci, Largo B. Pontecorvo, 3 - 56127 Pisa, Italy}
\author{J.W.~Mitchell}
\affiliation{Astroparticle Physics Laboratory, NASA/GSFC, Greenbelt, Maryland 20771, USA}
\author{S.~Miyake}
\affiliation{Department of Electrical and Electronic Systems Engineering, National Institute of Technology (KOSEN), Ibaraki College, 866 Nakane, Hitachinaka, Ibaraki 312-8508, Japan}
\author{A.A.~Moiseev}
\affiliation{Astroparticle Physics Laboratory, NASA/GSFC, Greenbelt, Maryland 20771, USA}
\affiliation{Center for Research and Exploration in Space Sciences and Technology, NASA/GSFC, Greenbelt, Maryland 20771, USA}
\affiliation{Department of Astronomy, University of Maryland, College Park, Maryland 20742, USA}\author{M.~Mori}
\affiliation{Department of Physical Sciences, College of Science and Engineering, Ritsumeikan University, Shiga 525-8577, Japan}
\author{N.~Mori}
\affiliation{INFN Sezione di Firenze, Via Sansone, 1 - 50019, Sesto Fiorentino, Italy}
\author{H.M.~Motz}
\affiliation{Faculty of Science and Engineering, Global Center for Science and Engineering, Waseda University, 3-4-1 Okubo, Shinjuku, Tokyo 169-8555, Japan}
\author{K.~Munakata}
\affiliation{Faculty of Science, Shinshu University, 3-1-1 Asahi, Matsumoto, Nagano 390-8621, Japan}
\author{S.~Nakahira}
\affiliation{Institute of Space and Astronautical Science, Japan Aerospace Exploration Agency, 3-1-1 Yoshinodai, Chuo, Sagamihara, Kanagawa 252-5210, Japan}
\author{J.~Nishimura}
\affiliation{Institute of Space and Astronautical Science, Japan Aerospace Exploration Agency, 3-1-1 Yoshinodai, Chuo, Sagamihara, Kanagawa 252-5210, Japan}
\author{S.~Okuno}
\affiliation{Kanagawa University, 3-27-1 Rokkakubashi, Kanagawa, Yokohama, Kanagawa 221-8686, Japan}
\author{J.F.~Ormes}
\affiliation{Department of Physics and Astronomy, University of Denver, Physics Building, Room 211, 2112 East Wesley Avenue, Denver, Colorado 80208-6900, USA}
\author{S.~Ozawa}
\affiliation{Quantum ICT Advanced Development Center, National Institute of Information and Communications Technology, 4-2-1 Nukui-Kitamachi, Koganei, Tokyo 184-8795, Japan}
\author{L.~Pacini}
\affiliation{INFN Sezione di Firenze, Via Sansone, 1 - 50019, Sesto Fiorentino, Italy}
\affiliation{Institute of Applied Physics (IFAC),  National Research Council (CNR), Via Madonna del Piano, 10, 50019, Sesto Fiorentino, Italy}
\author{P.~Papini}
\affiliation{INFN Sezione di Firenze, Via Sansone, 1 - 50019, Sesto Fiorentino, Italy}
\author{B.F.~Rauch}
\affiliation{Department of Physics and McDonnell Center for the Space Sciences, Washington University, One Brookings Drive, St. Louis, Missouri 63130-4899, USA}
\author{S.B.~Ricciarini}
\affiliation{INFN Sezione di Firenze, Via Sansone, 1 - 50019, Sesto Fiorentino, Italy}
\affiliation{Institute of Applied Physics (IFAC),  National Research Council (CNR), Via Madonna del Piano, 10, 50019, Sesto Fiorentino, Italy}\author{K.~Sakai}
\affiliation{Center for Space Sciences and Technology, University of Maryland, Baltimore County, 1000 Hilltop Circle, Baltimore, Maryland 21250, USA}
\affiliation{Astroparticle Physics Laboratory, NASA/GSFC, Greenbelt, Maryland 20771, USA}
\affiliation{Center for Research and Exploration in Space Sciences and Technology, NASA/GSFC, Greenbelt, Maryland 20771, USA}
\author{T.~Sakamoto}
\affiliation{College of Science and Engineering, Department of Physics and Mathematics, Aoyama Gakuin University,  5-10-1 Fuchinobe, Chuo, Sagamihara, Kanagawa 252-5258, Japan}
\author{M.~Sasaki}
\affiliation{Astroparticle Physics Laboratory, NASA/GSFC, Greenbelt, Maryland 20771, USA}
\affiliation{Center for Research and Exploration in Space Sciences and Technology, NASA/GSFC, Greenbelt, Maryland 20771, USA}
\affiliation{Department of Astronomy, University of Maryland, College Park, Maryland 20742, USA}\author{Y.~Shimizu}
\affiliation{Kanagawa University, 3-27-1 Rokkakubashi, Kanagawa, Yokohama, Kanagawa 221-8686, Japan}
\author{A.~Shiomi}
\affiliation{College of Industrial Technology, Nihon University, 1-2-1 Izumi, Narashino, Chiba 275-8575, Japan}
\author{P.~Spillantini}
\affiliation{Department of Physics, University of Florence, Via Sansone, 1 - 50019, Sesto Fiorentino, Italy}
\author{F.~Stolzi}
\affiliation{Department of Physical Sciences, Earth and Environment, University of Siena, via Roma 56, 53100 Siena, Italy}
\affiliation{INFN Sezione di Pisa, Polo Fibonacci, Largo B. Pontecorvo, 3 - 56127 Pisa, Italy}
\author{S.~Sugita}
\affiliation{College of Science and Engineering, Department of Physics and Mathematics, Aoyama Gakuin University,  5-10-1 Fuchinobe, Chuo, Sagamihara, Kanagawa 252-5258, Japan}
\author{A.~Sulaj} 
\affiliation{Department of Physical Sciences, Earth and Environment, University of Siena, via Roma 56, 53100 Siena, Italy}
\affiliation{INFN Sezione di Pisa, Polo Fibonacci, Largo B. Pontecorvo, 3 - 56127 Pisa, Italy}
\author{M.~Takita}
\affiliation{Institute for Cosmic Ray Research, The University of Tokyo, 5-1-5 Kashiwa-no-Ha, Kashiwa, Chiba 277-8582, Japan}
\author{T.~Tamura}
\affiliation{Kanagawa University, 3-27-1 Rokkakubashi, Kanagawa, Yokohama, Kanagawa 221-8686, Japan}
\author{T.~Terasawa}
\affiliation{Institute for Cosmic Ray Research, The University of Tokyo, 5-1-5 Kashiwa-no-Ha, Kashiwa, Chiba 277-8582, Japan}
\author{S.~Torii}
%\email[]{torii.shoji@waseda.jp}
\affiliation{Waseda Research Institute for Science and Engineering, Waseda University, 17 Kikuicho,  Shinjuku, Tokyo 162-0044, Japan}
\author{Y.~Tsunesada}
\affiliation{Graduate School of Science, Osaka Metropolitan University, Sugimoto, Sumiyoshi, Osaka 558-8585, Japan }
\affiliation{Nambu Yoichiro Institute for Theoretical and Experimental Physics, Osaka Metropolitan University,  Sugimoto, Sumiyoshi, Osaka  558-8585, Japan}
\author{Y.~Uchihori}
\affiliation{National Institutes for Quantum and Radiation Science and Technology, 4-9-1 Anagawa, Inage, Chiba 263-8555, Japan}
\author{E.~Vannuccini}
\affiliation{INFN Sezione di Firenze, Via Sansone, 1 - 50019, Sesto Fiorentino, Italy}
\author{J.P.~Wefel}
\affiliation{Department of Physics and Astronomy, Louisiana State University, 202 Nicholson Hall, Baton Rouge, Louisiana 70803, USA}
\author{K.~Yamaoka}
\affiliation{Nagoya University, Furo, Chikusa, Nagoya 464-8601, Japan}
\author{S.~Yanagita}
\affiliation{College of Science, Ibaraki University, 2-1-1 Bunkyo, Mito, Ibaraki 310-8512, Japan}
\author{A.~Yoshida}
\affiliation{College of Science and Engineering, Department of Physics and Mathematics, Aoyama Gakuin University,  5-10-1 Fuchinobe, Chuo, Sagamihara, Kanagawa 252-5258, Japan}
\author{K.~Yoshida}
\affiliation{Department of Electronic Information Systems, Shibaura Institute of Technology, 307 Fukasaku, Minuma, Saitama 337-8570, Japan}
\author{W.~V.~Zober}
\affiliation{Department of Physics and McDonnell Center for the Space Sciences, Washington University, One Brookings Drive, St. Louis, Missouri 63130-4899, USA}

\collaboration{CALET Collaboration}

\date{\today}

\begin{abstract} 
We present the results of a direct measurement of the cosmic-ray helium spectrum with the CALET instrument in operation on the International Space Station since 2015. 
The observation period covered by this analysis spans from October 13, 2015 to April 30, 2022 (2392 days).
The very wide dynamic range of CALET allowed to collect helium data over a large energy interval, from $\sim$40~GeV to $\sim$250~TeV, for the first time with a single instrument in Low Earth Orbit. 
The measured spectrum shows evidence of a deviation of the flux from a single power-law by more than 8~$\sigma$ with a progressive spectral hardening from a few hundred GeV to a few tens of TeV. 
This result is consistent with the data reported by space instruments including PAMELA, AMS-02, DAMPE and balloon instruments including CREAM.
At higher energy we report the onset of a softening of the helium spectrum around 30 TeV (total kinetic energy). 
Though affected by large uncertainties in the highest energy bins, the observation of a flux reduction  turns out to be consistent with the most recent results of DAMPE.
A Double Broken Power Law (DBPL) is found to fit simultaneously both spectral features: the hardening (at lower energy) and the softening (at higher energy).
A measurement of the proton to helium flux ratio in the energy range from 60~GeV/n to about 60~TeV/n is also presented, using the CALET proton flux recently updated with higher statistics.
\end{abstract}

%\pacs{
%  %96.50.S-,%cosmic rays
%  98.70.Sa, %cosmic rays
%  96.50.sb, %composition energy spectra
%  %95.35.+d, %Dark matter
%  95.55.Vj, %cosmic ray detectors
%  %95.85.Ry,%cosmic rays
%  29.40.Vj, %calorimeters
%  07.05.Kf %Data analysis:
%}

\maketitle

\section{Introduction} 
The observation of spectral features departing from a single power-law in the energy spectra of cosmic-ray nuclei can provide additional insight into the general phenomenology of cosmic-ray (CR) acceleration and propagation in the Galaxy.
The deviations observed by several experiments~\cite{ATIC2-p,CREAM-nuclei,CREAM-hardening,CREAM-I,PAMELA,AMS02-p,AMS02-he,CREAM-III-pHe,AMS-02-carbon,AMS-02-boron,AMS-02-nitrogen,CALET-CO,DAMPE-he,CALET-PROTON,new-CALET-p} are not easily accommodated within the conventional models of galactic cosmic-ray acceleration and propagation.
These unexpected features have prompted new theoretical interpretations in terms of revised acceleration and propagation mechanisms, 
as well as the possible contribution of local sources in the injection spectra of galactic cosmic rays~\cite{IP2-Malkov-2012, LS1-Erlykin-2012, LS2-Thoudam-2012, LS3-Bernard-2013, PR1-Blasi-2012, PR2-Aloisio-2013, RA2-Thoudam, SB2-Ohira-2011, SB3-Ohira-2016, SS1-Biermann-2010, SS2-Ptuskin-2013, SS3-Zatsepin-2006, Tomassetti-2012, Vladmirov-2012, Tomassetti2015, Evoli2018}.
Therefore, accurate measurements of the high-energy spectra of individual elements and of their flux ratios (most notably secondary-to-primary) are of particular interest to parameterize the energy dependence of spectral features in terms of spectral index variations and smoothness parameters.
Input from the new instruments launched to Low Earth Orbit in the last decade can provide additional discrimination power among the proposed theoretical models and improve our understanding of CR origin.
\\
\indent
At rigidities below a few TV, measurements are carried out either by magnetic spectrometers~\cite{PAMELA, AMS02-p} or calorimeters ~\cite{ATIC2-p,CREAM-I,CREAM-III-pHe,NUCLEON-JCAP,NUCLEON-JTEP}. The latter can reach a region of higher energies where new spectral features have been recently observed \cite{DAMPE-he,new-CALET-p,DAMPE-p}.
\\
\indent
The CALorimetric Electron Telescope (CALET)~\cite{Pier-ICRC21, Torii-ICRC19, torii2017,asaoka2018} is a space-based instrument equipped with a thick homogeneous calorimeter, optimized for the measurement of the all-electron spectrum~\cite{CALET2017,CALET2018}, yet with excellent capabilities to measure the hadronic component of cosmic rays including proton, light and heavy nuclei (up to nickel and above)~\cite{new-CALET-p,CALET-CO,CALET-Fe,CALET-Ni} in the energy range up to $\sim$1~PeV. 
In this Letter, we present a direct measurement of the cosmic-ray helium spectrum in kinetic energy $E$ from 40~GeV to 250~TeV with CALET.

\section{CALET Instrument}
CALET is an all-calorimetric instrument, consisting of three main sub-detectors. A charge detector (CHD) is followed by a 3 radiation-lengths ($X_0$) thick imaging calorimeter (IMC) and by a 27 $X_0$ thick total absorption calorimeter (TASC).
The CHD, positioned at the top of the apparatus, consists of a two layered hodoscope of plastic scintillators paddles, arranged along two orthogonal directions. 
The IMC is a fine grained sampling calorimeter alternating thin layers of Tungsten absorber with $x,y$ layers of scintillating fibers (with 1 mm$^2$ cross-section) read out individually. 
It reconstructs the early shower profile and the trajectory of the impinging particle with good angular resolution, also providing an independent charge measurement via multiple $dE$/$dx$ sampling \cite{pb2015}.
The TASC is an homogeneous calorimeter with 12 layers of tightly packed lead-tungstate (PbWO$_4$) logs, providing an energy measurement %providing a coarse imaging of the shower, and measuring its energy
over a very large dynamic range (more than 6 orders of magnitude) spanning four different gain ranges~\cite{asaoka2017}.
A more complete description of the instrument is given in the supplemental material of~\cite{CALET2017}.
\\
\indent
The instrument was launched on August 19, 2015 and emplaced on the JEM-EF (Japanese Experiment Module Exposed Facility) on the International Space Station, scientific observations~\cite{asaoka2018} started on October 13, 2015, and smooth and continuous operations have taken place since then.

\section{Data Analysis} 
Flight data collected from October 13, 2015 to April 30, 2022 were analyzed (2392 days). 
The total observation live time is 48459.7 hours and the live time fraction to total time is about 84.4\%."
The data analysis generally follows the same procedures used for the CALET analysis of protons~\cite{CALET-PROTON,new-CALET-p}, C-O~\cite{CALET-CO}, Fe~\cite{CALET-Fe} and Ni~\cite{CALET-Ni}.
\\
\indent
A highly efficient reconstruction of hadronic tracks is of primary importance for the flux measurement.
The Combinatorial Kalman Filter tracking algorithm (KF)~\cite{maestro2017}, already 
used in the proton spectrum analysis \cite{CALET-PROTON}, provides good performances also for helium tracks. 
\\
\indent
The shower energy of each event is calculated as the TASC energy deposit sum (hereafter $E_{\rm TASC}$), and is calibrated using penetrating protons and He particles selected in-flight by a dedicated trigger mode. 
A seamless stitching of adjacent gain ranges is performed on flight data and complemented by the confirmation of the instrument linearity over the whole range during pre-flight ground measurements with a UV pulsed laser, as described in Ref.~\cite{asaoka2017}.
\\
\indent
Time-dependent variations occurring during the long-term observation period are also corrected for each sensor, using penetrating particles as gain monitor~\cite{CALET2017}.
\\
\indent
Detailed Monte Carlo (MC) simulations have been performed, based on the EPICS simulation package \cite{EPICS,EPICSver}.
In order to assess the relatively large uncertainties in the modeling of hadronic interactions, a series of beam tests were carried out at the CERN-SPS using the CALET beam test model~\cite{akaike2013, niita2015, akaike2015}.
Trigger efficiency and energy response derived from MC simulations were tuned using the beam test results obtained in 2015 with ion beams of 13, 19 and 150 GeV/n. 
For helium nuclei a shower energy correction of 10.4\% (8\%) at 13 (19) GeV/n was applied, while a 3.2\% energy independent correction was applied at 150~GeV/n and above. A log-linear interpolation provided the correction factors for intermediate energies not measured at CERN.
No correction is applied to the trigger efficiency since beam test measurements are consistent with the MC simulations.
\\
\indent
In the analysis of hadrons, especially in the high-energy region where no beam test calibrations are possible, a comparison between different MC models is mandatory.
To this extent, we have run simulations with FLUKA~\cite{FLUKA-1,FLUKA-2,FLUKAver} and compared them with EPICS.
\\
\indent
A preselection of well-reconstructed and well-contained events is applied, prior to charge identification, to minimize the background contamination of the selected helium sample. The following criteria are applied.
\begin{enumerate*}[label=(\roman*),font={\bfseries}]
\item[Trigger:] only events taken with the on-board high-energy (HE) trigger mode are retained. This mode is designed to ensure maximum exposure to electrons above 10 GeV and to other high-energy shower events. 
Consistency between MC and flight data (FD) for triggered events is obtained by applying
an offline trigger filter requiring more severe conditions than the on-board trigger.
It removes residual effects due to positional and temporal variations of the detector gain.
\item[Track quality cut:]
selected events are required to have a good primary track candidate reconstructed in both views with the KF algorithm. A minimum number of points are required for each track segment and a $\chi^2$ cut is applied.
In this way an angular resolution for He nuclei of about 0.1$^{\circ}$ 
and an Impact Point (IP) resolution of $\sim 400\ \mu$m on the CHD top layer are achieved.
\item[Geometrical condition:] 
the reconstructed events are required to traverse the whole detector (i.e., from CHD top to TASC bottom, with 2 cm clearance from the edges of the TASC top layer) and be contained inside a fiducial region (acceptance A1), with a Geometric Factor (GF) of 0.051 m$^2$sr ($\sim$ 49\% of the total GF).
\item[Electron rejection:]
an electron rejection cut is applied, based on a fractional quantity known as ``Moli\`ere concentration along the track" and calculated by summing all energy deposits inside one Moli\`ere radius around each IMC fiber matched to the track and normalized to the total energy deposit sum in the IMC.
By requiring this quantity to be less than 0.75, when the fraction of the TASC energy deposited in the last layer is greater than 0.01,
more than 90\% of electrons are rejected while retaining a very high efficiency for helium nuclei ($>$ 99.9\% for $E > 50$ GeV). 
\item[Off-Acceptance rejection cuts:] hadronic interactions and the combinatorial track reconstruction are responsible for the occasional misidentification of one of the secondary tracks as the primary track. 
This results in a number of events erroneously reconstructed inside the fiducial acceptance A1.
To reject most of these events, different topological cuts are applied using the TASC information.
The fractional energy deposit in each of the first two TASC layers is required to be less than 0.3 to reject laterally incident tracks.
The residual between the impact points of a track onto the first two layers of the TASC and the center of gravity of the corresponding energy deposits is required (consistency cut) not to exceed the size of two PWO logs ($\pm 2$ cm). Taking advantage of the TASC granularity, the shower axis is reconstructed with the method of moments (see \cite{gomez1987} for details), and is required to cross the TASC-X1 layer. This cut rejects, with very high efficiency, lateral events erroneously reconstructed inside the fiducial region. A small correction (of a few \%) is applied to the cut efficiency to take into account small discrepancies between FD and MC.
\end{enumerate*}

The identification of cosmic-ray nuclei via a measurement of their charge is carried out with two independent subsystems that are routinely used to cross-calibrate each other: the CHD and the IMC. 
Tracking allows to select the CHD paddles crossed by the primary particle and, after application of position and time-dependent calibrations and corrections \cite{asaoka2017}, 
the information from the two CHD layers is combined into a single charge estimator.
The IMC, being equipped with individually readout scintillating fibers, provides multiple $dE/dx$ measurements up to a maximum of 16 samples.
The Interaction Point (IP) of the impinging particle is reconstructed at first~\cite{pb2015} and only the $dE/dx$ ionization clusters from the layers upstream the IP are used. 
The charge is evaluated as the truncated-mean of the valid samples rejecting 30\% of the highest ones. 
The non-linear response due to the saturation of the scintillation light in the fibers is corrected for, both in IMC and CHD, by fitting the light yield according to a quenching model described in Refs.~\cite{voltz,tarle}.
\\
\indent
To mitigate the effects of the increase of the backscattered background with energy, both charge measurements are calibrated to the nominal peak positions. 
This calibration is applied separately to FD and MC simulations by EPICS and FLUKA. 
To ensure a perfect match between FD and MC, the MC data are finely tuned with FD (separately for EPICS and FLUKA), fitting the proton and helium charge distributions in several energy slices with an asymmetric Landau distribution convoluted with a Gaussian.
The Full Width at Half Maximum (FWHM) and peak position of the charge distribution are extracted for each energy slice and used, on an event by event basis, to finely tune the MC distributions and to perform an energy dependent charge cut, resulting in an almost flat charge selection efficiency ($\sim$ 65\%). More details are given in the Supplemental Material ~\cite{thisSM}.
\\
\indent
Background contamination is estimated from MC simulations of protons, helium and from FD, as a function of the observed energy. 
The MC simulations are used to evaluate the relative contributions and the FD to assess the proton and helium relative abundances.
Charge contamination from protons misidentified as helium is the dominant component. Other not negligible contributions come from off-acceptance helium and protons mis-reconstructed inside the acceptance A1.
Depending on the energy, the estimated overall contamination ranges from a few percent to $\sim$20\% at the highest energies where the proton background becomes dominant.
The estimated background is then subtracted bin-by-bin from the $dN/dE$ distribution of helium candidates.
\\
\indent
In order to take into account the relatively limited energy resolution (the observed energy fraction is around 35\% and the energy resolution is 30\%--40\%), energy unfolding is necessary to correct for significant bin-to-bin migration effects and to infer the primary particle energy. 
In this analysis, we applied an iterative unfolding method based on the Bayes theorem \cite{dagost} implemented in the RooUnfold package \cite{roounf} in ROOT~\cite{root}, using the response matrix derived from the MC.
Convergence is obtained within two iterations, given the relatively accurate prior distribution obtained from the previous observations of AMS-02~\cite{AMS02-he} and CREAM-I~\cite{CREAM-I}. 
The energy bin width is chosen to be commensurate with the resolution of the TASC.
\\
\indent
The energy spectrum is obtained from the unfolded energy distribution as follows:
\begin{equation}
\Phi(E) = \dfrac{N(E)}{ \Delta E \times \varepsilon ( E ) \times S\Omega \times T}
\label{eq_flux}
\end{equation}
\begin{equation}
N(E) = U \left[N_{obs}(E_{\rm TASC}) - N_{bg}(E_{\rm TASC}) \right]
\end{equation}
where $\Delta E$ denotes the energy bin width, $E$ is the particle kinetic energy, calculated as the geometric mean of the lower and upper bounds of the bin, $N(E)$ is the bin content in the unfolded distribution, $\varepsilon (E)$ the overall selection efficiency (Fig. S2 of the SM \cite{thisSM}), $T$ is the live time, $S\Omega$ the ``fiducial'' geometrical acceptance, $U$  the unfolding procedure, $N_{obs}(E_{\rm TASC})$ the bin content of the observed energy distribution (including background), $N_{bg}(E_{\rm TASC})$ the background events in the same bin.

\section{Systematic Uncertainties}
The systematic uncertainties can be categorized into energy independent and energy dependent ones. The former includes systematic effects in the normalization and were studied in Ref.~\cite{CALET2017}.
This uncertainty is estimated around 4.1\% as the quadratic sum of the uncertainties on live time (3.4\%), radiation environment (1.8\%), and long-term stability (1.4\%). 
\\
The energy dependent uncertainties include the following contributions.
{\bf Trigger:}
the absolute calibration of the trigger efficiency was performed at the beam test. The main source of uncertainty comes from the accuracy of the calibration. 
A possible systematic bias in the trigger efficiency due to normalization was included in the uncertainty, 
by scanning the offline trigger threshold applied to TASCX1 signal between 100 and 150 MIP signal. 
{\bf Shower energy correction:}
the absolute calibration of the energy response in the low-energy region was carried out using the beam test data. 
Both the accuracy of the calibration and the uncertainty in the model used to fit the test beam data are taken into account in the systematics. 
{\bf Track reconstruction and acceptance:}
the effects of tracking on the flux were evaluated by studying its dependence on the goodness-of-tracking cuts. 
To investigate the uncertainty on the acceptance, restricted acceptance regions have been studied and the resultant fluxes were compared. 
{\bf Background subtraction:}
background subtraction is only slightly dependent on the simulated spectral shape.  
Different re-weighting functions (including $E^\alpha$ with $-2.9 \leq \alpha \leq -2.5$) were adopted for the MC spectrum and the relative differences with respect to the reference case were included in the systematic uncertainty for each energy bin. 
{\bf Unfolding:} 
the uncertainties from the unfolding procedure were evaluated by applying different response matrices computed by varying the spectral index (between $-2.9$ and $-2.5$) of the MC generation spectrum, or the number of iterations of the Bayesian method. 
{\bf Charge ID and Off-Acceptance Rejection cuts:}
the flux stability against the selection cut efficiencies was studied 
around the reference value and the differences with respect to the reference case were accounted as systematic error.
The thresholds of each cut were varied separately in an appropriate range ($\pm 1$ FWHM for the charge ID cut) around the reference value and the differences versus the reference case were accounted as systematic error.
{\bf MC model:}
a second Monte Carlo (FLUKA) is used to evaluate the smearing matrix and the relevant selection efficiencies. 
For each bin, a systematic error is obtained by a comparison of FLUKA with EPICS results. 
\\
\indent
Considering all of the above contributions, the total systematic uncertainty remains below 10\% up to $\sim$60 TeV.
Above it increases moderately, remaining commensurate with the statistical error as summarized in Fig.~S5 of the SM~\cite{thisSM} where the total uncertainty is shown with all the relevant contributions listed above.
\\
\indent
Two independent helium analyses were carried out by separate groups inside the CALET collaboration, using different event selections and background rejection procedures. 
The results of the two analysis are consistent with each other within the errors.

\begin{figure}[bth]
\begin{center}
%\vspace{0.025cm}
\includegraphics[width=\hsize]{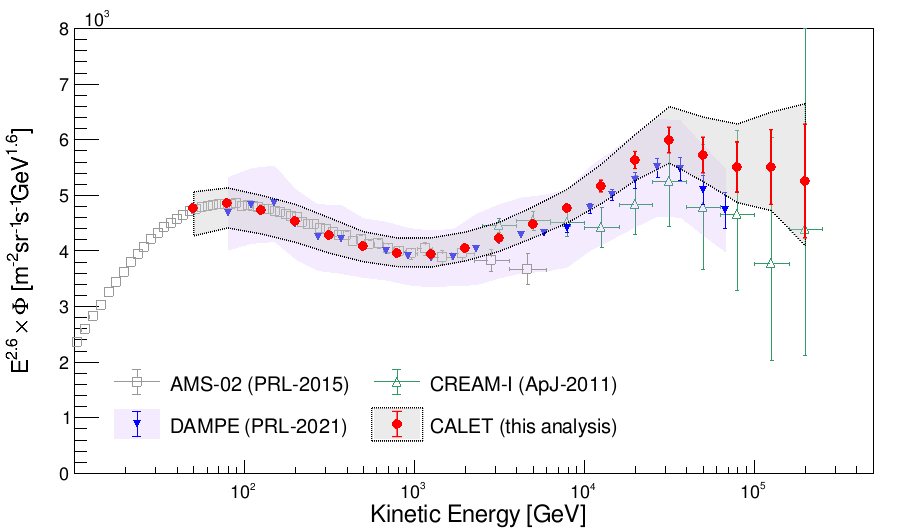}
\caption{Cosmic ray helium spectrum measured by CALET (red markers), compared with previous direct observations~\cite{AMS02-he,DAMPE-he,CREAM-I}. 
The error bars represent only the statistical error, the gray band represents the quadratic sum of statistical and systematic error.
The light violet colored band show the systematic uncertainty of Ref.~\cite{DAMPE-he}.
}
\label{fig:Heflux}
\end{center}
\end{figure}
\section{Results} 
The energy spectrum of CR helium, as measured by CALET in an interval of kinetic energy per particle from $\sim$40 GeV to $\sim$250 TeV, is shown in Fig. \ref{fig:Heflux} where the statistical and systematic uncertainties are bounded within a gray band.
The measured helium flux and the statistical and systematic errors are tabulated in Table I of the SM~\cite{thisSM}. 
The CALET spectrum is compared with previous observations from space-based \cite{AMS02-he,DAMPE-he} and balloon-borne \cite{CREAM-I,CREAM-III-pHe} experiments.
Our spectrum is in good agreement with the very accurate measurements by AMS-02 in the lower energy region below a few TeV, as well as with the measurements from calorimetric instruments in the higher energy region, in particular with the recent measurement of DAMPE~\cite{DAMPE-he}.
\begin{figure}[bth!]
\begin{center}
\vspace{0.2cm}
\includegraphics[width=\hsize]{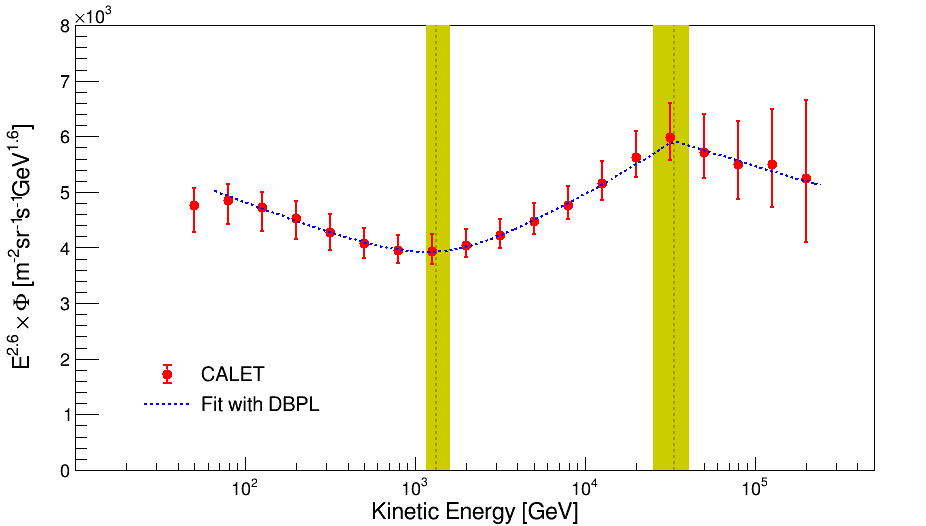} 
\caption{
Fit of CALET data with a DBPL function (\ref{eq_fit}). The result is consistent with other recent measurements~\cite{DAMPE-he} within the errors. Both statistical and systematic uncertainties are taken into account \cite{thisSM}. 
}
\label{fig:fitDBPL}
\end{center}
\end{figure}
\begin{figure}[bth!]
\begin{center}
\includegraphics[width=\hsize]{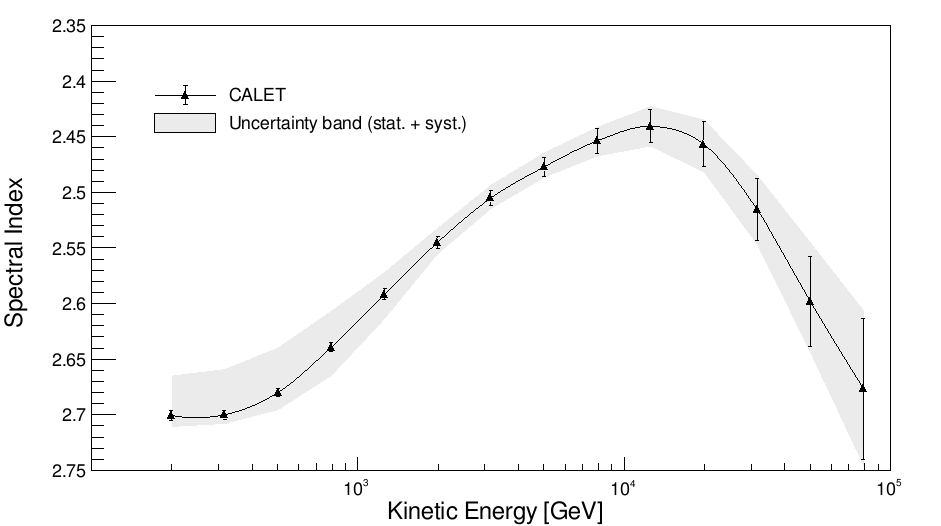} 
\caption{
Energy dependence of the spectral index calculated within a sliding energy window for CALET data. The spectral index is determined for each point by fitting the data using $\pm 2$ bins. The gray band indicates the uncertainty range including systematics.
}
\label{fig:fitSW}
\end{center}
\end{figure}

In Fig. \ref{fig:fitDBPL}, a fit of CALET data with a ``Double Broken Power-Law'' (DBPL), Eq.~\ref{eq_fit}, is shown in the energy range from 60 GeV to 250 TeV:
\begin{equation}
\resizebox{.9\hsize}{!}{
$\Phi(E) = C \left( \frac{E}{GeV} \right)^\gamma \left[1 + \left(\frac{E}{E_0}\right)^S \right]^{\frac{\Delta\gamma}{S}} \left[1 + \left(\frac{E}{E_1}\right)^{S_1} \right]^{\frac{\Delta\gamma_1}{S_1}}$
}
\label{eq_fit}
\end{equation}
A progressive hardening from a few hundred GeV to a few tens TeV is observed. The fit returns a power law index 
$\gamma = -2.703\ ^{+0.005}_{-0.006}\ (stat)\ ^{+0.032}_{-0.009}\ (syst)$, 
$\Delta \gamma  = 0.25\ ^{+0.02}_{-0.01}\ (stat)\ ^{+0.02}_{-0.03}\ (syst)$, first break energy 
$E_{0} = 1319\ ^{+113}_{-93}\ (stat)\ ^{+267}_{-124}\ (syst)$ GeV and smoothness parameter 
$S = 2.7\ ^{+0.6}_{-0.5}\ (stat)\ ^{+3.0}_{-0.9}\ (syst)$.
The onset of a flux softening above a few tens of TeV is also observed, with a second spectral index variation 
$\Delta\gamma_{1} = -0.22\ ^{+0.07}_{-0.10}\ (stat)\ ^{+0.03}_{-0.04}\ (syst)$ and
second break energy $E_{1} = 33.2 \ ^{+9.8}_{-6.2}\ (stat)\ ^{+1.8}_{-2.3}\ (syst)$ TeV. 
Given the relatively large uncertainties of the data in the highest energy bins, the second smoothness parameter $S_{1}$ cannot be effectively constrained and is kept fixed at value $S_{1} = 30$. 
\\
\indent
The index change $\Delta \gamma$ is proven to be different from zero by more than 8 $\sigma$, taking into account both statistical and systematic error~\cite{thisSM}. 
The fit parameters are generally consistent, within the errors, with the recent results of DAMPE \cite{DAMPE-he}, although $\Delta\gamma_{1}$ seems to indicate a less pronounced softening in our data.
\\
\indent
The spectral hardening and softening can be easily seen in Fig.~\ref{fig:fitSW} where the spectral index is shown as a function of kinetic energy.
For each point the spectral index is fitted within a sliding energy window of $\pm 2$ bins. The black marker in the plot represents the index $\gamma$ with its statistical error, while the gray band represents the quadratic sum of statistical end systematic uncertainties.
\begin{figure}[bth!]
\begin{center}
\includegraphics[width=\hsize]{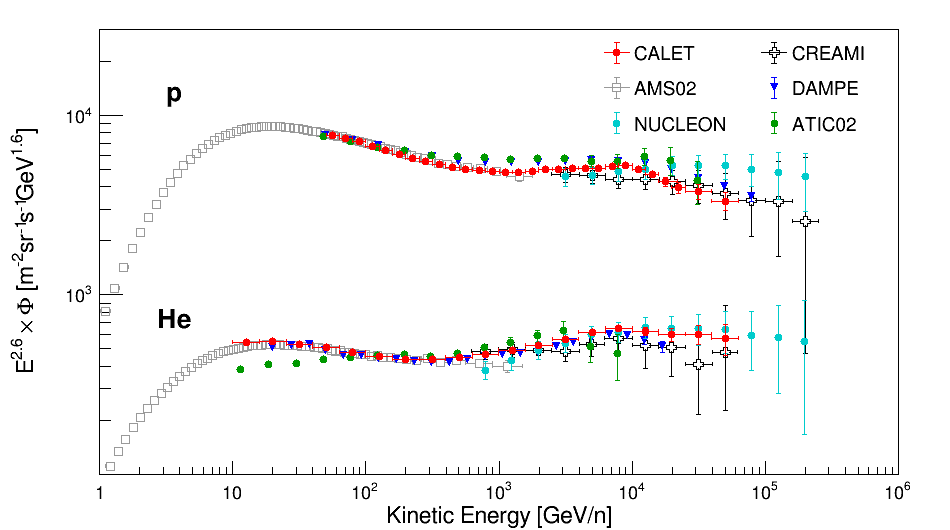}
\caption{
The CALET proton~\cite{new-CALET-p} and helium fluxes are shown as function of kinetic energy per nucleon, together with previous measurements from other experiments~\cite{AMS02-p,AMS02-he,DAMPE-p,DAMPE-he,CREAM-I,NUCLEON-pHe,ATIC2}.
}
\label{fig:CpHe}
\end{center}
\end{figure}
\begin{figure}[bth!]
\begin{center}
\includegraphics[width=\hsize]{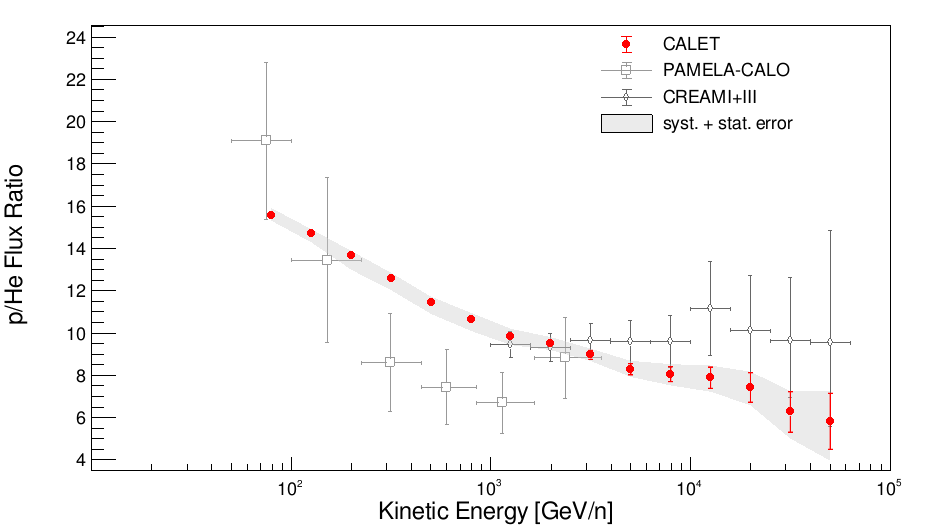} 
\caption{
Energy spectrum of p/He ratio as measured by CALET, the red bars represent statistical error only, the gray band represents the quadratic sum of statistical and systematic errors. 
Results of previous measurements from CREAM~\cite{CREAM-III-pHe} and PAMELA (calorimeter analysis)~\cite{PAMELA-CALO,CRDB} are shown as reference.
}
\label{fig:CpHeR}
\end{center}
\end{figure}

Differences between the proton and helium spectra can contribute important constraints on acceleration models (e.g.~\cite{IP2-Malkov-2012}).
To ease the comparison in Fig.~\ref{fig:CpHe}, we show the CALET proton spectrum published in Ref.~\cite{new-CALET-p} and the helium spectrum from this analysis, in kinetic energy per nucleon. 
The $^3$He contribution to the flux is taken into account assuming the same $^3$He/$^4$He ratio as measured by AMS-02 \cite{AMS02-3he4he} and extrapolating it to higher energies with use of a single power-law fit.\\ 

Using the CALET proton flux of Ref.~\cite{new-CALET-p}, we present the p/He flux ratio in Fig.~\ref{fig:CpHeR} as measured by CALET with high statistical precision in a wide energy range from 60 GeV/n to $\sim$60 TeV/n. 
Both the statistical and systematic errors are shown; details on the systematic uncertainty can be found in the SM~\cite{thisSM}. 
Measurements from other experiments~\cite{CREAM-III-pHe,PAMELA-CALO} are included in the same plot.
Our result is found to be in agreement with previous measurements from magnetic spectrometers~\cite{AMS02-he,PAMELA} up to their maximum detectable rigidity ($\sim$2 TV), 
as shown in Fig. S8 of the SM~\cite{thisSM}.
The measured p/He ratio is tabulated in Table II and III of the SM~\cite{thisSM}, as a function of kinetic energy per nucleon and rigidity respectively.

\section{Conclusion} 
CALET has measured the cosmic-ray helium spectrum covering, for the first time with a single instrument on the ISS, the large energy range from 40~GeV to 250~TeV. 
Our spectrum is not consistent with a single power law (at $>$ 8~$\sigma$ level) and its shape confirms the presence of a hardening above a few hundred GeV (where a SBPL function fits the spectrum well) and the onset of a flux softening above a few tens TeV. 
A DBPL fits both spectral features with parameters that are found to be consistent, within the errors, with the most recent results of DAMPE \cite{DAMPE-he}.
Using the CALET proton flux~\cite{new-CALET-p}, we also measured the p/He ratio in the interval from 60 GeV/n to $\sim$60 TeV/n. Due to the partial cancellation of systematic errors in the ratio, this measurement can provide important information on the respective acceleration and propagation mechanisms.

\section{Acknowledgments} 
\begin{acknowledgments}
We gratefully acknowledge JAXA’s contributions to the development of CALET and to the operations aboard the JEM-EF on the International Space Station.
We also wish to express our sincere gratitude to ASI (Agenzia Spaziale Italiana) and NASA for their support of the CALET project. 
This work was supported in part by JSPS Grant-in-Aid for Scientific Research (S) Grant No. 19H05608, and by the MEXT-Supported Program for the Strategic Research Foundation at Private Universities (2011–2015)(Grant No. S1101021) at Waseda University.
The CALET effort in Italy is supported by ASI under agreement 2013-018-R.0 and its amendments. 
The CALET effort in the United States is supported by NASA through Grants No. 80NSSC20K0397, No. 80NSSC20K0399, No. NNH18ZDA001N-APRA18-0004, and under award number 80GSFC21M0002.
\end{acknowledgments}

%%%%%%%%%%%%%%%%%%%%%%%%%%%%
\providecommand{\noopsort}[1]{}\providecommand{\singleletter}[1]{#1}%
%

%%%%%%%%%% Merge with supplemental materials %%%%%%%%%%
%\pagebreak
%%%%%%%%%%%%%%%%%%%%%%%%%%%%%%%%%%%%%%%%%%%%%%%%%%%%%%%
%%%%%%%%%%%%%%%%%%%%%%%%%%%%%%%%%%%%%%%%%%%%%%%%%%%%%%%
%%%%%%%%%%%%%%%%%%%%%%%%%%%%%%%%%%%%%%%%%%%%%%%%%%%%%%%
%%%%%%%%%%%%%%%%%%%%%%%%%%%%%%%%%%%%%%%%%%%%%%%%%%%%%%
\widetext
\clearpage
%%%%%%%%%% Merge with supplemental materials %%%%%%%%%%
\setcounter{equation}{0}
\setcounter{figure}{0}
\setcounter{table}{0}
\setcounter{page}{1}
\makeatletter
\renewcommand{\theequation}{S\arabic{equation}}
\renewcommand{\thefigure}{S\arabic{figure}}
\renewcommand{\bibnumfmt}[1]{[S#1]}
\renewcommand{\citenumfont}[1]{S#1}
%%%%%%%%%%%%%%%%%%%%%%%%%%%%%%%%%%%%%%%%%%%%%%%%%%%%%%%
\begin{center}
\textbf{\Large Direct Measurement of the Cosmic-Ray Helium Spectrum\\ 
 from 40~GeV to 250~TeV with the Calorimetric Electron Telescope\\
 on the International Space Station\\
 \vspace*{0.2cm}
SUPPLEMENTAL MATERIAL}\\
\vspace*{0.2cm}
(CALET collaboration)

\end{center}
\vspace*{0.5cm}
Supplemental material concerning ``Direct Measurement of the Cosmic-Ray Helium Spectrum from 40~GeV to 250~TeV with the Calorimetric Electron Telescope on the International Space Station.''
\vspace*{1cm}

%%%%%%%%%%%%%%%%%
\clearpage
%%%%%%%%%%%%%%%%%%%%%%%%%%%%%%%%%%
\section{CALET Helium candidate}
Figure \ref{fig:CALETevt} shows an example of helium candidate in CALET. 
The display is representative of a typical well reconstructed helium nucleus crossing all sub-detectors.
The selected event has a shower energy of about 700 GeV in the TASC.
The blue lines represent the projections of the reconstructed impinging particle trajectory in the $X-Z$ and $Y-Z$ views respectively.
\begin{figure}[bth!]
\begin{center}
\includegraphics[width=.85\textwidth]{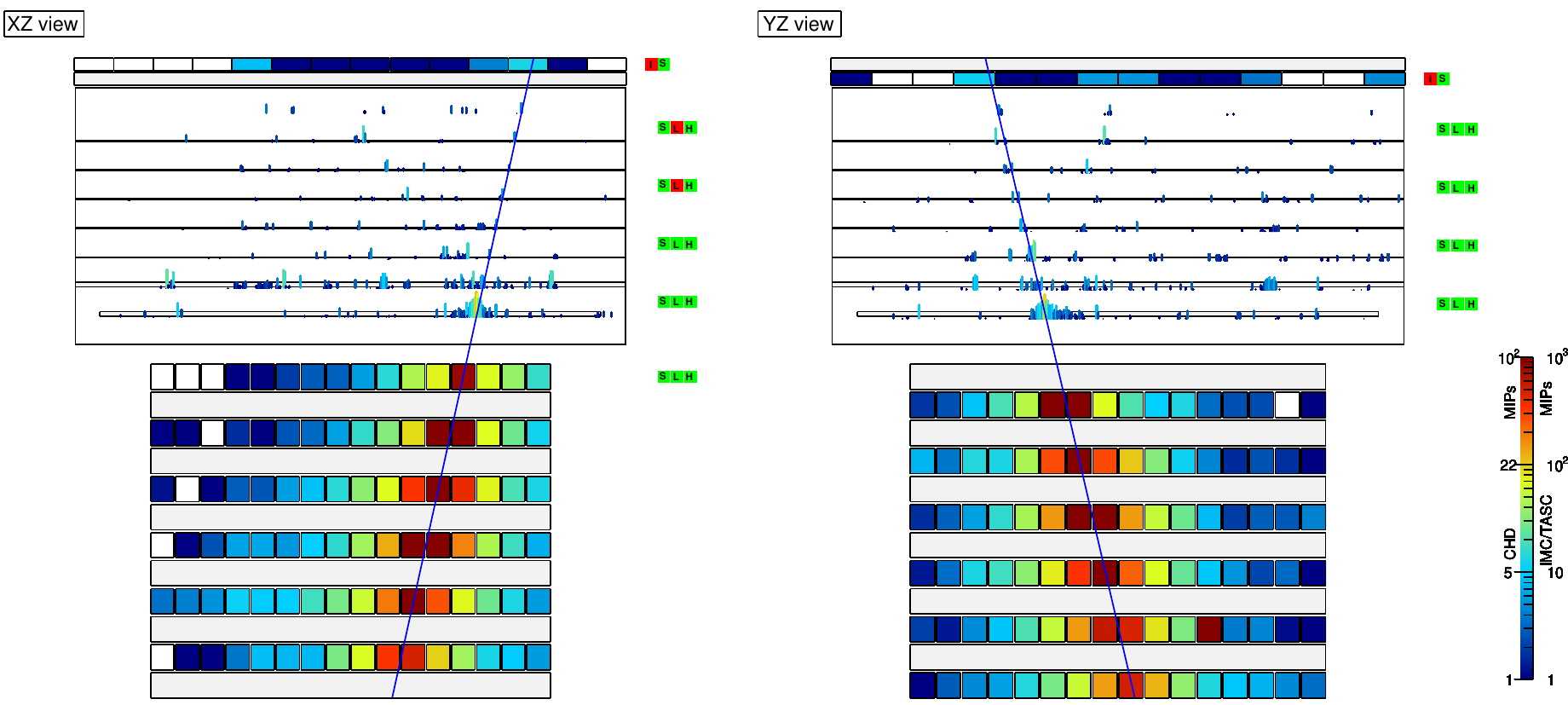}%width=.95\textwidth
\caption{Event display of a characteristic helium candidate.}
\label{fig:CALETevt}
\end{center}
\end{figure}
\clearpage

\section{CALET selection efficiency}
Figure \ref{fig:CALETeff} shows the total selection efficiency for helium nuclei (blue squares) estimated for CALET with EPICS MC simulations.
In the same plot the charge selection (magenta triangles) and the HE trigger (black circles) relative efficiencies are shown, representing the two main contributions to the overall selection efficiency.
\begin{figure}[h!]
\begin{center}
\includegraphics[width=.85\textwidth]{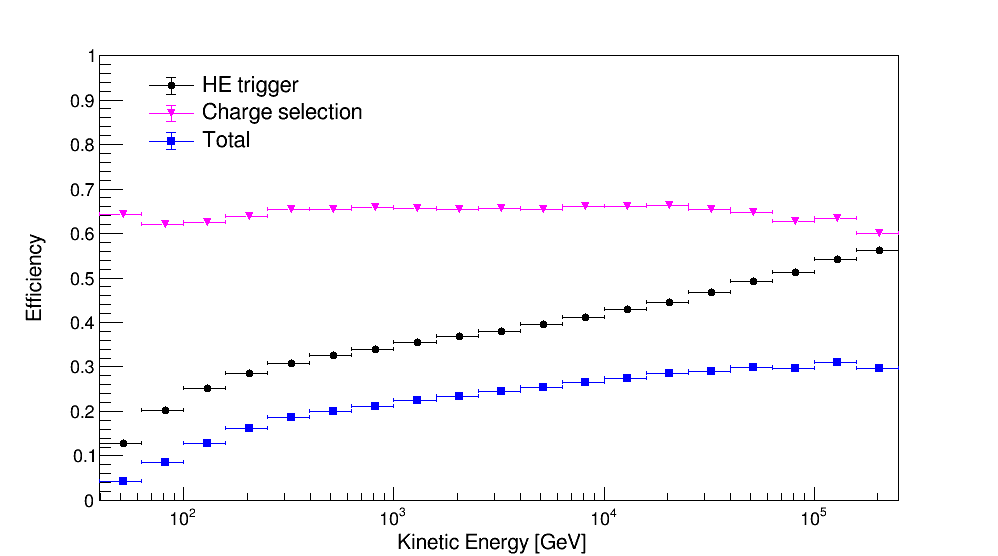}
\caption{The total selection efficiency for helium nuclei (blue squares) is shown together with the charge selection (magenta triangles) and the HE trigger (black circles) relative efficiencies.
}
\label{fig:CALETeff}
\end{center}
\end{figure}
\clearpage

\section{Charge calibration and identification}
Both CHD and IMC charge measurements are calibrated and corrected for their non-linear response due to the saturation of the scintillation light, and for the energy shift related to the backscattering background increasing with energy.\\
In order to have a perfect match between FD and MC, the MC data are fine tuned to the flight data \cite{ICRC21-pb}.
This additional calibration is performed fitting proton and helium charge distributions in several energy intervals (hereafter referred to as slices) with an asymmetric Landau distribution convoluted with a Gaussian (see the left panel of figure \ref{fig:CALETchID} for an example).
Then, the FWHM and peak position of the charge distribution are computed for each energy slice, together with the Left and Right handed Half-Width-at-Half-Maximum (LWHM, RWHM), and fitted to the whole energy range with a logarithmic polynomial (dashed lines in the right panel of figure \ref{fig:CALETchID}). 
The fits of the peak position and FWHM values are used, on an event by event basis, to fine tune the MC distributions. 
The fits to LWHM and RWHM values are used to perform an energy dependent charge cut to select the helium candidates, by applying simultaneous window cuts on the CHD and IMC reconstructed charges, requiring $3 \times \text{LWHM} < Z_{CHD} < 5 \times \text{RWHM}$ and $3\times \text{LWHM} < Z_{IMC} < 5 \times \text{RWHM}$.
An almost flat charge selection efficiency (close to 65\%) is obtained, as shown in figure \ref{fig:CALETeff} by the magenta triangle-shaped markers. 
\begin{figure}[h!]
\begin{center}
\includegraphics[width=.9\textwidth]{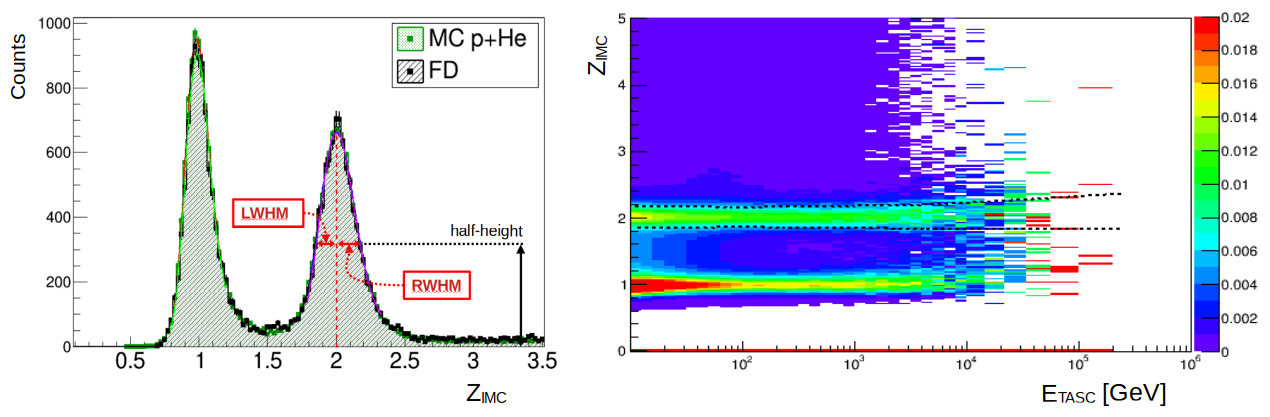}
\caption{ Left panel: the IMC charge distribution in a given energy slice for FD (black) and the  sum of proton and helium MC (green). ``Lan-gauss'' fit to the charge peak distribution is shown as an example, where the left and right width at half maximum of the curve are also shown.
Right panel: IMC charge distribution dependence on the TASC shower energy deposit, the black dashed line shows the width of the distribution.
}
\label{fig:CALETchID}
\end{center}
\end{figure}
\clearpage

\subsection{Energy Unfolding}
In order to account for bin-to-bin migration effects due to the limited energy resolution, energy unfolding is applied to correct the  $E_{\rm TASC}$ distribution of the selected Helium candidates and to infer the primary particle energy. 
In this analysis, we apply the iterative unfolding method based on the Bayes theorem \cite{dagost-S} implemented in the RooUnfold package \cite{roounf-S,root-S}. 
Figure \ref{fig:CALETsmatrix} shows the response matrix used in the unfolding procedure, which is derived using the EPICS MC simulation and applying the same selection as for FD. 
Each element of the matrix represents the probability that a primary helium nucleus in a given energy interval produces energy deposits in multiple bins of $E_{\rm TASC}$.
\begin{figure}[h!]
\begin{center}
\includegraphics[width=.6\textwidth]{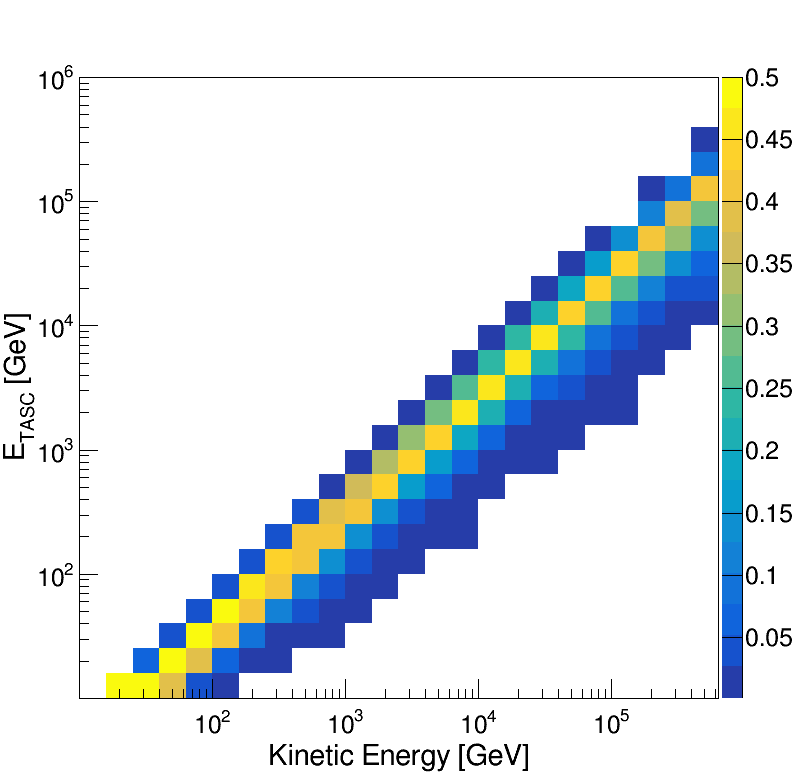}
\caption{Response matrix for helium, derived from MC simulations of CALET with EPICS, by applying the same selection as for FD. 
The color scale is associated with the probability that nuclei in a given bin of kinetic energy generate showers in different intervals of deposited energy in the TASC ($E_{\rm TASC}$).}
\label{fig:CALETsmatrix}
\end{center}
\end{figure}
\clearpage

\section{Systematic Uncertainties}%
A detailed breakdown of the systematic uncertainties in the helium flux measurement is shown in figure \ref{fig:CALETsys}, where each line and error bar represents the contribution of a different source of systematic error to the total uncertainty which is calculated as the sum in quadrature of all the known contributions and is represented by the band within the green lines in the top panel of the figure \ref{fig:CALETsys}.
On the bottom panel, the teal filled band represents the energy independent contribution of the systematic error, while all the other colored lines and bars show the individual energy dependent contributions.
They include: charge identification (cyan dot-dashed lines), off-acceptance rejection cuts (black lines), geometrical acceptance and track quality cuts (magenta lines), offline trigger (azure dashed line), MC model (yellow bars), shower energy correction (blue lines), energy unfolding (gray lines) and background subtraction (dark green bars).
\\ \indent
The systematic uncertainty of fit parameters are evaluated as follows. All the spectra used for the estimate of each source of systematic uncertainties (i.e. charge, trigger, etc.), that are obtained by varying the thresholds of the relevant cuts and the analysis parameters, are fitted with a DBPL function (Eq. 3 in the main body of the paper). Then, for each fit, the maximum difference (with either sign) between the obtained parameters and the one of the reference spectrum is taken as an estimate of the systematic error related to that source. The total uncertainty for each parameter is therefore obtained as the quadratic sum of the errors related to each systematic source.
For the index change parameter ($\Delta\gamma$), the sum in quadrature of the total systematic uncertainty and the statistical error proves the $\Delta\gamma$ to be different from zero by more than 8 $\sigma$.
\begin{figure}[h!]
\begin{center}
\includegraphics[width=\hsize]{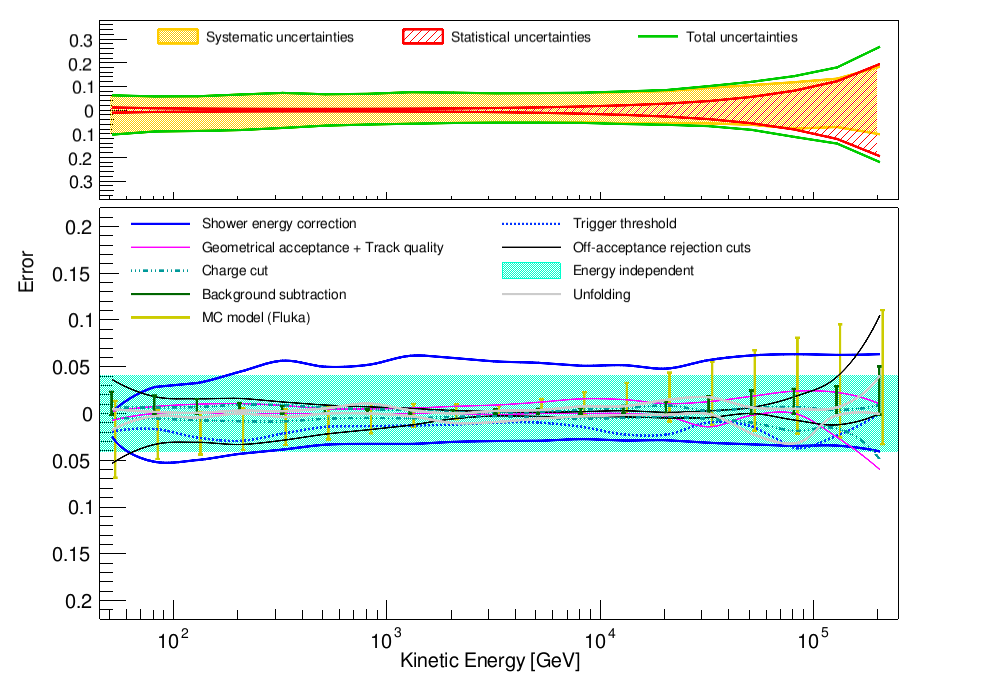}%width=.95\textwidth
\caption{ 
Energy dependence (vs particle kinetic energy expressed in GeV) of systematic uncertainties (relative errors) for helium nuclei. 
On the top panel the band within the green lines shows the sum in quadrature of all the sources of systematic uncertainties. 
On the bottom panel the teal colored band represents the energy independent contribution of systematic error in each energy bin. 
All the other lines and bars show the detailed breakdown of systematic errors, stemming from charge identification, off-acceptance rejection cuts, geometrical acceptance and track quality cuts, offline trigger, MC model, shower energy correction, energy unfolding and background subtraction.
}
\label{fig:CALETsys}
\end{center}
\end{figure}
\clearpage

Figure \ref{fig:CALETsysRatio} shows a detailed breakdown of the systematic uncertainties relative to the proton/helium flux ratio, where each line represents the contribution of a different source of systematic error to the total uncertainty, calculated as the sum in quadrature of all the contributions and represented by the band within the green lines in the top panel of the same figure.
On the bottom panel, the colored lines show the individual contributions of: charge identification (magenta), off-acceptance rejection cuts (cyan), MC model (blue dashed) and energy unfolding (black dotted).
\\ \indent
The systematic uncertainty in the p/He ratio is evaluated considering both the systematic errors of the helium flux (reported above) and of the proton flux, as reported in~\cite{new-CALET-p-S}. 
For each relevant source of systematic uncertainty two different p/He ratios have been determined by calculating the fluxes at both sides of the relative error bands. The relative differences of these ratios with respect to the reference case were accounted for as systematic error.
Since the proton and helium fluxes are measured with the same detector, the shower energy correction, the trigger threshold, the geometrical acceptance and the energy independent systematic are expected to give similar contributions to the two fluxes and therefore be suppressed in the ratio.
\begin{figure}[h!]
\begin{center}
\includegraphics[width=\hsize]{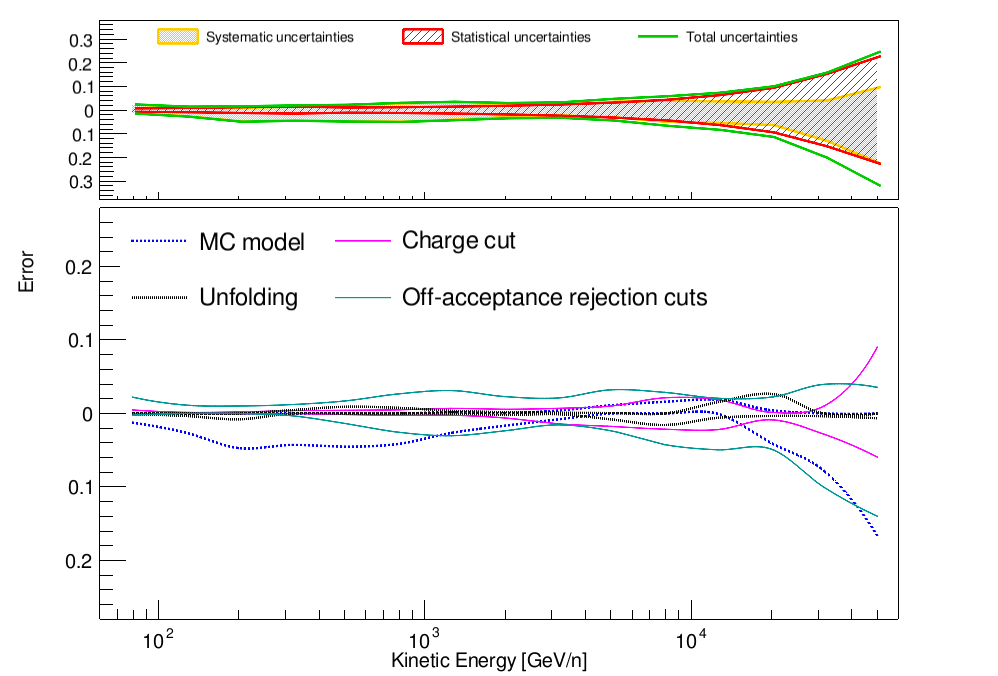}
\caption{ 
Energy dependence (vs kinetic energy per nucleon expressed in GeV/n) of systematic uncertainties (relative errors) for proton/helium ratio. 
On the top panel the band within the green lines shows the sum in quadrature of all the sources of systematic uncertainties. 
On the bottom panel the lines show the detailed breakdown of systematic errors, stemming from charge identification (magenta), off-acceptance rejection cuts (cyan), MC model (blue dashed) and energy unfolding (black dotted).
}
\label{fig:CALETsysRatio}
\end{center}
\end{figure}
\clearpage

\section{Results}
Figure \ref{fig:CALETflux} shows an enlarged version of Fig. 1 in the main body of the paper, where the data from Refs \cite{NUCLEON-pHe-S,ATIC2-S} are added to the comparison.
The energy spectrum of CR helium, as measured by CALET in an interval of kinetic energy per particle from $\sim$40 GeV to $\sim$250 TeV is presented.
The red markers represent the statistical errors, while the gray band is bound by the quadratic sum of statistical and systematic errors.

The bottom panel of Fig. \ref{fig:CpHeR_S} shows the p/He flux ratio measured by CALET as a function of rigidity. 
The red bars represent the statistical errors and the gray band represents the quadratic sum of statistical and systematic errors.
The CALET result is found to be in agreement with previous measurements from the magnetic spectrometers AMS-02~\cite{AMS02-he-S} and PAMELA~\cite{PAMELA-S}, shown in the same plot as reference. 
For the sake of completeness, in the top panel of the same figure the CALET proton~\cite{new-CALET-p-S} and helium fluxes are shown as a function of rigidity, together with previous measurements from other experiments~\cite{AMS02-p-S,AMS-02-carbon-S,PAMELA-S}.
\begin{figure}[h!]
\begin{center}
\includegraphics[width=\hsize]{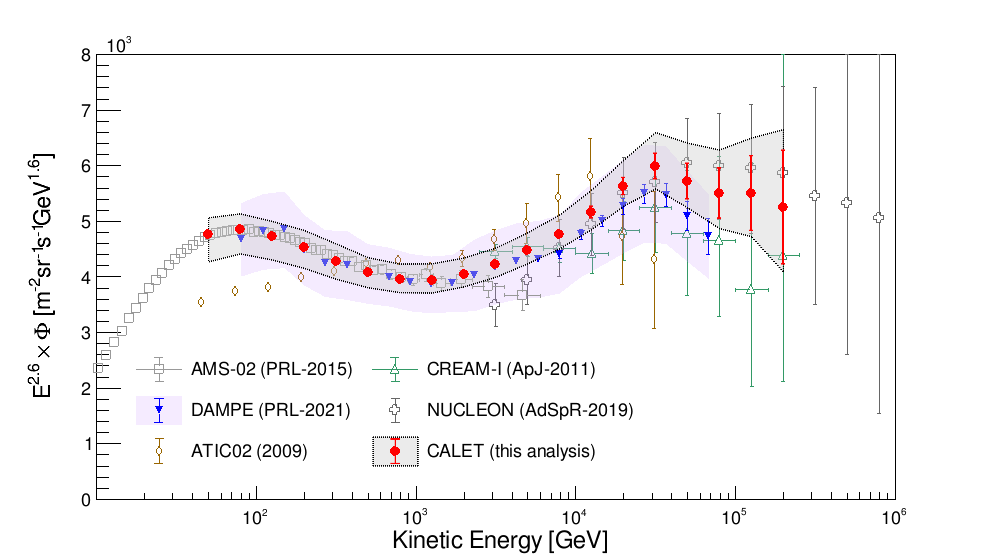}
\caption{Enlarged version of the cosmic-ray helium spectrum measured by CALET (red markers), compared with previous direct observations \cite{NUCLEON-pHe-S,ATIC2-S,AMS02-he-S,DAMPE-he-S,CREAM-I-S}. %\cite{AMS02-he,DAMPE-he,CREAM-I}. 
The CALET error bars refers to statistical error only, the gray band represents the quadratic sum of statistical and systematic errors.
The blue bars and the light violet colored band show the statistical and systematic uncertainty for DAMPE~\cite{DAMPE-he-S}, respectively.
For ATIC02~\cite{ATIC2-S} only the statistical error is represented.
}
\label{fig:CALETflux}
\end{center}
\end{figure}
\begin{figure}[bth!]
\begin{center}
\includegraphics[width=\hsize]{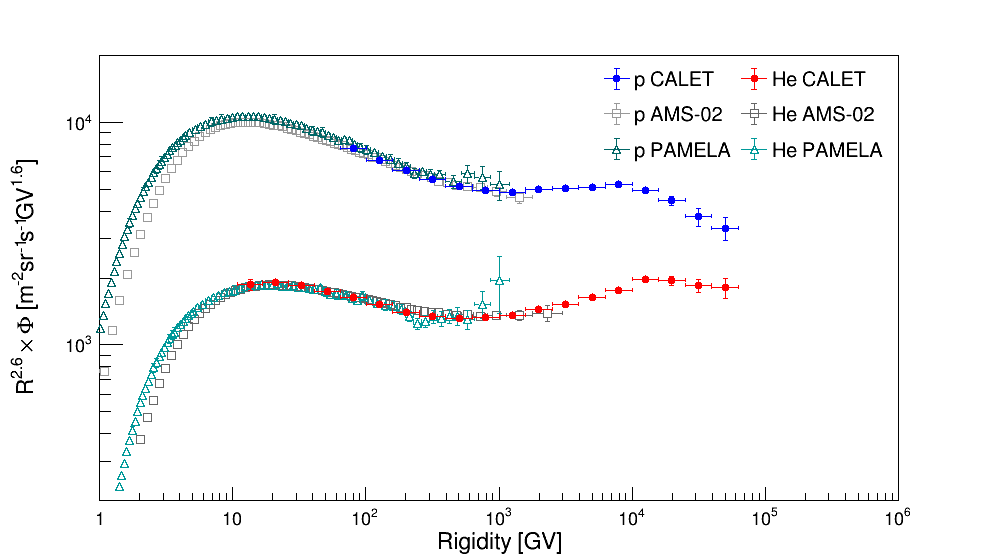} 
\includegraphics[width=\hsize]{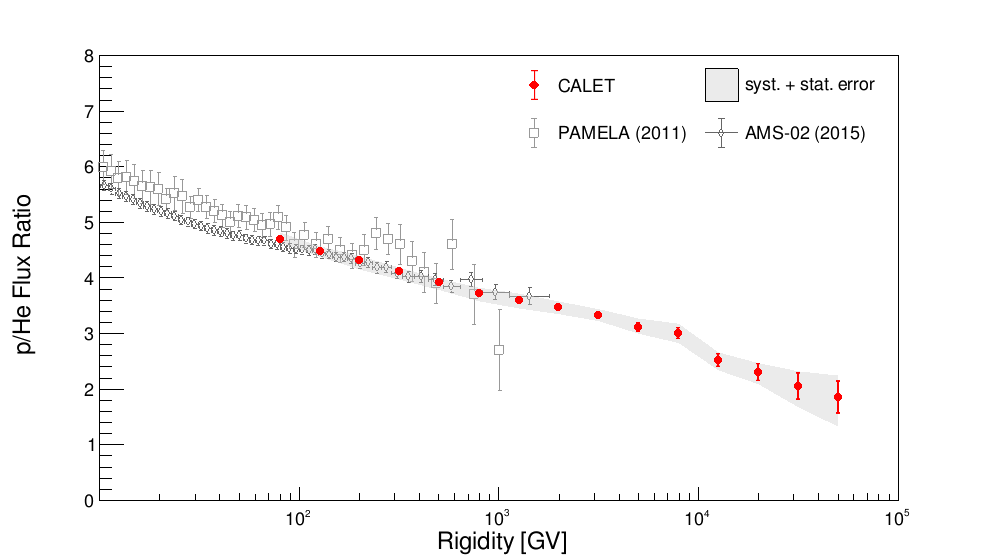} 
\caption{
The top panel shows the CALET proton~\cite{new-CALET-p-S} and helium fluxes as a function of rigidity, together with previous measurements from other experiments~\cite{AMS02-p-S,AMS-02-carbon-S,PAMELA-S}, only the statistical errors are shown.
The bottom panel shows the p/He ratio as measured by CALET as a function of rigidity, the red vertical bars represent statistical error only while the gray band represents the quadratic sum of statistical and systematic errors. 
Previous measurements from AMS-02~\cite{AMS02-he-S} and PAMELA~\cite{PAMELA-S} are shown as reference.
}
\label{fig:CpHeR_S}
\end{center}
\end{figure}
\clearpage

%%%%%%%%%%%%%%%%%%%%%%%%%%%%%%%%%%%%%%

\clearpage
\renewcommand{\arraystretch}{1.25}
\begin{table*}
  \caption{Table of CALET helium differential spectrum. 
  The first, second and third error on the flux represents the statistical uncertainty, systematic error in normalization, and energy dependent systematic uncertainties, respectively.
\label{tab:Heflux}}
\begin{ruledtabular}
\begin{tabular}{c c c}
Energy Bin [GeV] & Flux [m$^{-2}$sr$^{-1}$s$^{-1}$(GeV)$^{-1}$]   \\
\hline
39.8--63.1 & $(1.812 \, \pm 0.020\, \pm 0.074 \, _{-0.171}^{+0.084}) \times 10^{-1}$ \\
63.1--100.0 & $(5.573 \, \pm 0.044\, \pm 0.228 \, _{-0.448}^{+0.220}) \times 10^{-2}$ \\
100.0--158.5 & $(1.637 \, \pm 0.010\, \pm 0.067 \, _{-0.128}^{+0.066}) \times 10^{-2}$ \\
158.5--251.2 & $(4.752 \, \pm 0.026\, \pm 0.195 \, _{-0.350}^{+0.246}) \times 10^{-3}$ \\
251.2--398.1 & $(1.354 \, \pm 0.007\, \pm 0.056 \, _{-0.088}^{+0.081}) \times 10^{-3}$ \\
398.1--631.0 & $(3.890 \, \pm 0.020\, \pm 0.160 \, _{-0.199}^{+0.206}) \times 10^{-4}$ \\
631.0--1000.0 & $(1.142 \, \pm 0.006\, \pm 0.047 \, _{-0.051}^{+0.063}) \times 10^{-4}$ \\
1000.0--1584.9 & $(3.430 \, \pm 0.020\, \pm 0.141 \, _{-0.138}^{+0.217}) \times 10^{-5}$ \\
1584.9--2511.9 & $(1.063 \, \pm 0.007\, \pm 0.044 \, _{-0.038}^{+0.065}) \times 10^{-5}$ \\
2511.9--3981.1 & $(3.352 \, \pm 0.030\, \pm 0.137 \, _{-0.109}^{+0.192}) \times 10^{-6}$ \\
3981.1--6309.6 & $(1.074 \, \pm 0.012\, \pm 0.044 \, _{-0.035}^{+0.063}) \times 10^{-6}$ \\
6309.6--10000.0 & $(3.452 \, \pm 0.052\, \pm 0.142 \, _{-0.113}^{+0.206}) \times 10^{-7}$ \\
10000.0--15848.9 & $(1.128 \, \pm 0.023\, \pm 0.046 \, _{-0.043}^{+0.072}) \times 10^{-7}$ \\
15848.9--25118.8 & $(3.714 \, \pm 0.102\, \pm 0.152 \, _{-0.175}^{+0.256}) \times 10^{-8}$ \\
25118.8--39810.7 & $(1.195 \, \pm 0.046\, \pm 0.049 \, _{-0.065}^{+0.101}) \times 10^{-8}$ \\
39810.7--63095.6 & $(3.449 \, \pm 0.192\, \pm 0.141 \, _{-0.161}^{+0.351}) \times 10^{-9}$ \\
63095.6--100000.0 & $(9.997 \, \pm 0.821\, \pm 0.410 \, _{-0.667}^{+1.453}) \times 10^{-10}$ \\
100000.0--158489.1 & $(3.025 \, \pm 0.369\, \pm 0.124 \, _{-0.162}^{+0.478}) \times 10^{-10}$ \\
158489.1--251188.4 & $(8.707 \, \pm 1.695\, \pm 0.357 \, _{-0.814}^{+1.651}) \times 10^{-11}$ \\
\end{tabular}
\end{ruledtabular}
\end{table*}
\renewcommand{\arraystretch}{1.0}

%\clearpage
\renewcommand{\arraystretch}{1.25}
\begin{table*}
  \caption{Table of CALET proton to helium flux ratio in kinetic energy per nucleon (GeV$/n$).
  The first and second error on the ratio represents the statistical uncertainty and the systematic uncertainties, respectively.
\label{tab:pHeratio}}
\begin{ruledtabular}
\begin{tabular}{c c c}
Energy Bin [GeV$/n$] & Ratio \\
\hline
63.1--100.0 & $15.5 \, \pm 0.1 \, _{-0.2}^{+0.4}$ \\
100.0--158.5 & $14.7 \, \pm 0.1 \, _{-0.4}^{+0.2}$ \\
158.5--251.2 & $13.7 \, \pm 0.2 \, _{-0.7}^{+0.1}$ \\
251.2--398.1 & $12.6 \, \pm 0.2 \, _{-0.5}^{+0.2}$ \\
398.1--631.0 & $11.5 \, \pm 0.1 \, _{-0.5}^{+0.2}$ \\
631.0--1000.0 & $10.6 \, \pm 0.1 \, _{-0.5}^{+0.3}$ \\
1000.0--1584.9 & $9.8 \, \pm 0.1 \, _{-0.4}^{+0.3}$ \\
1584.9--2511.9 & $9.5 \, \pm 0.2 \, _{-0.3}^{+0.2}$ \\
2511.9--3981.1 & $9.0 \, \pm 0.2 \, _{-0.2}^{+0.2}$ \\
3981.1--6309.6 & $8.3 \, \pm 0.3 \, _{-0.3}^{+0.3}$ \\
6309.6--10000.0 & $8.1 \, \pm 0.4 \, _{-0.4}^{+0.3}$ \\
10000.0--15848.9 & $7.9 \, \pm 0.5 \, _{-0.4}^{+0.3}$ \\
15848.9--25118.8 & $7.4 \, \pm 0.7 \, _{-0.5}^{+0.3}$ \\
25118.8--39810.7 & $6.3 \, \pm 1.0 \, _{-0.8}^{+0.3}$ \\
39810.7--63095.6 & $5.8 \, \pm 1.3 \, _{-1.3}^{+0.6}$ \\
\end{tabular}
\end{ruledtabular}
\end{table*}
\renewcommand{\arraystretch}{1.0}

\clearpage
\renewcommand{\arraystretch}{1.25}
\begin{table*}
  \caption{Table of CALET proton to helium flux ratio in rigidity (GV). The error in the ratio refers to the statistical uncertainty only.
  The first and second error represents the statistical uncertainties and the systematic uncertainties, respectively.
}
\begin{ruledtabular}
\begin{tabular}{c c c}
Rigidity Bin [GV] & Ratio \\
\hline
64.0--100.9 & $4.69 \, \pm 0.04 \, _{-0.04}^{+0.12}$ \\
100.9--159.4 & $4.47 \, \pm 0.04 \, _{-0.08}^{+0.10}$ \\
159.4--252.1 & $4.32 \, \pm 0.05 \, _{-0.16}^{+0.05}$ \\
252.1--399.0 & $4.12 \, \pm 0.06 \, _{-0.13}^{+0.04}$ \\
399.0--631.9 & $3.92 \, \pm 0.04 \, _{-0.13}^{+0.05}$ \\
631.9--1000.9 & $3.72 \, \pm 0.04 \, _{-0.14}^{+0.09}$ \\
1000.9--1585.8 & $3.59 \, \pm 0.04 \, _{-0.14}^{+0.12}$ \\
1585.8--2512.8 & $3.47 \, \pm 0.05 \, _{-0.11}^{+0.10}$ \\
2512.8--3982.0 & $3.32 \, \pm 0.06 \, _{-0.08}^{+0.10}$ \\
3982.0--6310.5 & $3.12 \, \pm 0.08 \, _{-0.09}^{+0.14}$ \\
6310.5--10000.9 & $3.01 \, \pm 0.10 \, _{-0.15}^{+0.14}$ \\
10000.9--15849.8 & $2.52 \, \pm 0.11 \, _{-0.14}^{+0.09}$ \\
15849.8--25119.8 & $2.30 \, \pm 0.15 \, _{-0.14}^{+0.08}$ \\
25119.8--39811.6 & $2.04 \, \pm 0.24 \, _{-0.29}^{+0.15}$ \\
39811.6--63096.6 & $1.85 \, \pm 0.29 \, _{-0.44}^{+0.28}$ \\
\end{tabular}
\end{ruledtabular}
\end{table*}
\renewcommand{\arraystretch}{1.0}

\providecommand{\noopsort}[1]{}\providecommand{\singleletter}[1]{#1}%


\begin{thebibliography}{62}
\bibitem [{EPI()}]{EPICSver}   
	\BibitemOpen  \href@noop {} 
	{\bibinfo  {journal} {EPICS and COSMOS versions are 9.20 and  8.00, respectively}}
	\BibitemShut {NoStop}%
\end{thebibliography}
\end{document}